\documentclass[nofootinbib,aps,amssymb,floatfix,twocolumn,superscriptaddress,preprintnumbers]{revtex4}

\usepackage{graphicx}
\usepackage{bm}
\usepackage{amsmath}
\usepackage{amssymb}
\usepackage{amsfonts}
\usepackage{float}
\usepackage{hyperref}
\usepackage{dsfont}  
\usepackage{slashed}  
\usepackage{booktabs}
\usepackage{multirow}%https://www.overleaf.com/project/5f75d0158ffdb40001394bbb
\usepackage{subfigure}
\usepackage[sort&compress]{natbib}
\usepackage{xcolor}
\usepackage{ulem}

\newcommand{\be}{\begin{equation}}  
\newcommand{\ee}{\end{equation}}  
\newcommand{\beq}{\begin{eqnarray}} 
\newcommand{\eeq}{\end{eqnarray}}

\newcommand{\bea}{\begin{eqnarray}}
\newcommand{\eea}{\end{eqnarray}}

\newcommand{\nn}{\nonumber \\}

\begin{document}

\title{Spin-spin entanglement in diffractive heavy quark production}

\author{Michael Fucilla}
\affiliation{ National Centre for Nuclear Research, Pasteura 7, Warsaw 02-093, Poland}

\author{Yoshitaka Hatta  }
%\email{yhatta@bnl.gov}
\affiliation{Physics Department, Brookhaven National Laboratory, Upton, NY 11973, USA}
\affiliation{RIKEN BNL Research Center, Brookhaven National Laboratory, Upton, NY 11973, USA}

\begin{abstract}
  We calculate the spin density matrix of a heavy quark-antiquark pair ($b\bar{b}$, $c\bar{c}$ or $s\bar{s}$)   diffractively produced in Deep Inelastic Scattering and  Ultraperipheral Collisions.  
  We show that the Pomeron exchange leaves  characteristic imprints on the entanglement pattern between the quark and the antiquark.
  For the  longitudinally polarized virtual photon, the pair always exhibits maximal entanglement and maximal violation of the Bell-CHSH  inequality. For the transversely polarized  photon, the pair is always entangled and Bell-violating, reaching maximal entanglement and maximal violation simultaneously  when the transverse momentum approximately equals the quark mass.

\end{abstract}

\maketitle

{\it Introduction}---The recent experimental measurements of the spin-spin  entanglement in top-antitop ($t\bar{t}$) quark  pairs by the ATLAS 
\cite{ATLAS:2023fsd} and the CMS \cite{CMS:2024pts} collaborations at the LHC have garnered a significant attention   amid a surge of interest in the intersection between collider physics and quantum information science \cite{Barr:2024djo,Afik:2022kwm,Afik:2025ejh}. While quantum entanglement among elementary particles such as electrons and photons has been studied since the early days of quantum mechanics \cite{Einstein:1935rr,Wu:1950zz,Bohm:1951xw,Bell:1964kc,Clauser:1969ny},  the direct observation of spin entanglement for quarks in QCD is highly nontrivial due to confinement. The top quarks are special because they decay  by the electroweak interaction $t\to b+W^+,\bar{t}\to \bar{b}+W^-$ before the strong interaction kicks in. The information about the spin state of the $t\bar{t}$ pair  can be recovered by measuring the angular distribution of the leptons $\ell^{\pm}$ produced    in the subsequent decays $W^+\to \ell^++\nu_\ell$, $W^-\to \ell^-+\bar{\nu}_{\ell}$ \cite{Baumgart:2012ay,Afik:2020onf,Aguilar-Saavedra:2023hss,Han:2023fci,Dong:2023xiw}.

Another, related paradigm is the test of Bell's inequality \cite{Bell:1964kc}, or more specifically, the Clauser-Horne-Shimony-Holt (CHSH) inequality \cite{Clauser:1969ny}. Bipartite states that violate the Bell-CHSH inequality are said to exhibit `Bell nonlocality', a phenomenon  that reflects more subtle and intricate quantum correlations than entanglement.  In general, Bell nonlocality means entanglement but the reverse is not  true. As a result, establishing Bell nonlocality usually requires much stricter experimental tests  \cite{Fabbrichesi:2021npl,Severi:2021cnj}, and has not yet been achieved  in the top quark sector.

These quantum phenomena of course exist also in lighter quark systems  such as bottom ($b\bar{b}$), charm ($c\bar{c}$) and strangeness ($s\bar{s}$) pairs. However, the corresponding measurements  are more challenging because the pair  first  fragments  hadronically. Fortunately, the spin state of the pair is partially retained when fragmenting into heavy baryons  $q\to \Lambda_q$ and $\bar{q}\to \bar{\Lambda}_q$ \cite{Galanti:2015pqa,Lin:2025eci}  which then decay semi-leptonically. (See also an alternative approach \cite{Cheng:2025cuv}.) Although such processes suffer from small branching ratios, measurements may still be possible \cite{Kats:2023zxb,Afik:2025grr} given the high luminosity of the LHC.  The recent $\Lambda\bar{\Lambda}$ measurement  at RHIC \cite{STAR:2025njp} is also encouraging.

It is natural to consider initiating  similar experimental programs in Deep Inelastic Scattering (DIS) at the Electron-Ion Collider (EIC) \cite{AbdulKhalek:2021gbh} that offers high luminosity and a cleaner experimental  environment due to lepton scattering.  While top quarks cannot be produced at the EIC, at least  $c\bar{c}$ and $s\bar{s}$ pairs are copiously produced, and $b\bar{b}$ events can also be measured depending on kinematics.  A related process is Ultraperipheral Collisions (UPC) \cite{Baltz:2007kq}   where a heavy nucleus acts as a source of real photons. This mimics the photo-production limit of DIS, and can be studied  already at the existing experimental facilities such as  RHIC and the LHC. The spin density matrix of the $q\bar{q}$ pair from the lowest order process $\gamma^*+g \to q+\bar{q}$  has been recently calculated  \cite{Qi:2025onf}. Remarkably, for the longitudinally polarized virtual photon $\gamma^*_L$,  the produced $q\bar{q}$ pair is found to be always maximally entangled, and at the same time, the Bell-CHSH inequality is maximally violated. 

The $2\to 2$ process $\gamma^*+g \to q+\bar{q}$  corresponds to inclusive production where the target proton/nucleus breaks up. At the EIC, one can also study the exclusive diffractive  production of $q\bar{q}$ pairs where the target stays intact. This arises from the underlying process $\gamma^*+{\mathbb P}\to q+\bar{q}$ where ${\mathbb P}$ denotes `Pomeron', the color-singlet gluonic exchange in the $t$-channel. Despite decades of intense scrutiny, the nature of the QCD Pomeron and the associated high density gluonic matter is not fully understood. In this paper, we investigate the quantum informational aspect of the Pomeron by calculating the spin density matrix in $\gamma^*+{\mathbb P}\to q+\bar{q}$. We will be particularly interested in to what extent the Pomeron entangles the $q\bar{q}$ pair in spin space and whether the pair observes the Bell-CHSH inequality. Our finding provides theoretical  motivation for a new interdisciplinary research direction  that can be explored  at the existing and future facilities.  For previous works  on different types of entanglement in the context of high energy QCD, see  \cite{Kovner:2015hga,Peschanski:2016hgk,Kharzeev:2017qzs,Liu:2018gae,Ramos:2020kyc,Bhattacharya:2024sno,Guo:2024jch,Brandenburg:2024ksp,Hatta:2024lbw,Dumitru:2025bib,Agrawal:2025yoe,Ouchen:2025tta,Hentschinski:2025pyq}.

{\it Heavy quark pair production in diffractive DIS}---Consider exclusive heavy quark pair production in unpolarized electron-proton (or electron-nucleus) scattering $e+p\to e'+\gamma^*+p \to e'+q+\bar{q}+p'$.

We have in mind bottom, charm and strange quarks $q=b,c,s$. First we work in a frame in which the virtual photon with momentum $q^\mu$ and virtuality $q^2=-Q^2$ moves fast  in the $+x^3$ direction and a $q\bar{q}$ pair is created in the photon  fragmentation region with momenta 
\beq
\tilde{k}^\mu = \left(zq^+, \frac{k_\perp^2+m^2}{2zq^+},{\bm k}\right), \ 
\tilde{k}'^\mu = \left(\bar{z}q^+,\frac{k_\perp^2+m^2}{2\bar{z} q^+},-{\bm k}\right) ,\label{fast} \nonumber
\eeq
where  $m$ is the quark mass. $z$ ($\bar{z}=1-z$) is the momentum fraction of the photon carried by the (anti)quark. It is related to the rapidities $y,\bar{y}$ as  $z=e^y/(e^y+e^{\bar{y}})$. 
The light-cone coordinates  $p^\mu=(p^+,p^-,{\bm p})$ are defined by $p^\pm=\frac{1}{\sqrt{2}}(p^0\pm p^3)$ and  two-dimensional transverse vectors are denoted by boldface letters ${\bm p}=(p^1,p^2)$ with $p_\perp \equiv |{\bm p}|$.  
The pair has invariant mass 
\beq
M^2= (\tilde{k}+\tilde{k}')^2= \frac{k_\perp^2+m^2}{z\bar{z}}. 
\eeq
For simplicity, the proton recoil momentum ${\bm \Delta}=-{\bm k} -{\bm k'}\approx 0$ has been neglected.   

We assume the `Regge' kinematics where the $\gamma^*+p$ center-of-mass (CM) energy $W^2=(p+q)^2$ is large.   In the eikonal approximation, for the longitudinally polarized virtual photon, the cross section is given by  \cite{Bartels:1996ne,Nikolaev:2003zf,Altinoluk:2015dpi,Hatta:2016dxp,Boussarie:2019ero}.
\beq
&&\hspace{-6mm}  \left.\frac{d \sigma^{ L}_{\alpha \alpha' \beta \beta'}}{d z  d^2 \boldsymbol{k} d^2{\bm \Delta} }\right|_{{\bm \Delta}\approx 0} =   \frac{  \alpha_{\rm em} e_q^2z \bar{z} Q^2}{ N_c (q^+)^2}   \bigg|  \int \frac{  d^2 \boldsymbol{p}T \left( {\bm p} \right) }{(\boldsymbol{k}-\boldsymbol{p})^2+\mu^2}  \bigg |^2  
\nn
&& \qquad \qquad \times 
\bar{v}_{\beta'} ( \tilde{k}' ) \gamma^+ u_{\beta} (\tilde{k}) \bar{u}_{\alpha} ( \tilde{k} ) \gamma^+ v_{\alpha'} ( \tilde{k}'), 
\label{lsigma}
\eeq
where $\mu^2\equiv z\bar{z}Q^2+m^2$ and  $e_q$ is the quark electromagnetic charge in units of $|e|$. $\alpha,\alpha'$ and $\beta,\beta'$ denote the spin states of the $q\bar{q}$ pair in the amplitude and the complex-conjugate amplitude, respectively. Since our goal is to compute the spin density matrix,  these indices have not been summed over \cite{Bernreuther:1993hq,Brandenburg:1998xw,Baumgart:2012ay}. The T-matrix of the $q\bar{q}$ pair (`color dipole') $T$  represents the Pomeron exchange in the present context.  Our normalization is  such that, in  
the Golec-Biernat-Wusthoff (GBW) model \cite{Golec-Biernat:1998zce},
\beq
T({\bm p})=\frac{N_c \sigma_0}{(2\pi)^2}\left(\delta^{(2)}({\bm p})- \frac{R^2}{\pi}e^{-R^2p^2_\perp}\right), \label{gbw}
\eeq
where $\sigma_0$ is the effective  transverse area of the proton and $R$ depends on $W$ and $Q$ \cite{Golec-Biernat:1998zce}. 
In the following we use the notations  
\beq
T_1\equiv \int  \frac{d^2 \boldsymbol{p}\, T({\bm p})}{({\bm k}-{\bm p})^2+\mu^2}, \quad 
T_2\equiv \frac{-1}{k_\perp^2}\int  \frac{d^2 \boldsymbol{p} \, {\bm k}\cdot {\bm p}T({\bm p})}{({\bm k}-{\bm p})^2+\mu^2}  . \label{t1t2}
\eeq

To evaluate (\ref{lsigma}), we first go to the pair's  CM frame via a Lorentz boost $\tilde{k}^\pm= e^{\pm\eta}k^\pm$, $\tilde{k}'^\pm= e^{\pm\eta}k'^\pm$ with the  boost factor  
$e^\eta
= q^+\sqrt{\frac{2z\bar{z}}{k_\perp^2+m^2}}$. 
In this frame, the quark has velocity 
\beq
|\vec{k}|=\frac{M}{2}\beta, \qquad \beta = \sqrt{1-\frac{4m^2}{M^2}}, \label{beta}
\eeq
and propagates in the direction
\beq
\cos\theta = \frac{k^3}{|\vec k|}  = \frac{(z-\bar{z})M}{\sqrt{M^2-4m^2}}  , \qquad \phi={\rm arg}(k^1+ik^2), \label{theta}
\eeq
in the polar coordinates measured from the $+x^3$ axis. 
The resulting spinor bilinear $\bar{u}(k)\gamma^+ v(k')$ is most conveniently analyzed in the coordinate system spanned by the orthonormal basis  $\{\hat{n},\hat{r},\hat{k}\}$ \cite{Baumgart:2012ay,Afik:2020onf} where  $\hat{k}=(\sin\theta \cos \phi, \sin \theta \sin \phi, \cos \theta)$ and   
\beq
\hat{n} = \frac{\hat{x}^3\times \hat{k}}{\sin \theta} ,
%= (-\sin \phi, \cos\phi,0), 
\qquad 
\hat{r}= \frac{\hat{x}^3-\hat{k}\cos\theta}{\sin\theta} . 
\eeq
We can then use the helicity basis  along the $\hat{k}$ direction \cite{Jacob:1959at} and express (\ref{lsigma}) in the form 
\beq
 \frac{d \sigma_{\alpha \alpha' \beta \beta'}^{ L}}{d z  d^2 \boldsymbol{k} d^2{\bm \Delta} } =\frac{A^L}{4} \left(\delta_{\alpha\beta}\delta_{\alpha'\beta'}  +C_{ab}^L \xi_\alpha^\dagger \sigma^a \xi_\beta \eta_{\alpha'}^\dagger \sigma^b \eta_{\beta'} \right) , \label{rep}
\eeq 
where $a,b=n,r,k$, and $\sigma^a$ are the Pauli matrices. $\xi$, $\eta$ are the two-component quark and antiquark spinors, respectively. We find 
\begin{equation}
    A^L  %= \Gamma_L \frac{M^2}{4} (1-\beta^2 \cos^2{\theta}) 
    = \frac{8\alpha_{em}e_q^2z^2\bar{z}^2Q^2}{N_c} T_1^2,
\end{equation}
and 
\beq
&&C_{nn}^L = 1, \nn
&&C_{rr}^L=-C_{kk}^L = -\frac{1-(2-\beta^2)\cos^2\theta}{1-\beta^2\cos^2\theta},\nn 
&& C_{rk}^L=C_{kr}^L = -\frac{\sqrt{1-\beta^2}\sin 2\theta}{1-\beta^2\cos^2\theta}.
\label{cij}
\eeq
Note that $(C_{rr}^L)^2+(C_{rk}^L)^2=1$. 
The sign convention of $C_{ab}$ is  such that the matrix $C_{ab}$ represents  the correlation between the spin projections of the quark and the antiquark along the $+\hat{k}$ axis. For the antiquark   moving in the $-\hat{k}$ direction, this is  opposite to the helicity. 
It turns out that the matrix elements $C^L_{ab}$ do not depend on $Q$ and are exactly the same as for the `one-gluon exchange' process  $\gamma^*_L+g\to q+\bar{q}$ \cite{Qi:2025onf}. This is somewhat surprising because the Pomeron  exchanged in the $t$-channel consists of an arbitrary number of gluons ${\mathbb P}=gg,ggg,\cdots$ in an overall color singlet state. It also means that $C_{ab}^L$ carries no information about the property of the target proton.   As observed in 
\cite{Qi:2025onf}, the states represented by the density matrix (\ref{cij}) are maximally entangled pure states.  In the relativistic limit $\beta \to 1$ and/or for the symmetric configuration  $z=\frac{1}{2}$ (meaning $\cos\theta=0$ in the CM frame), it reduces to one of the Bell states.

Another measure of quantum correlation is the violation of the Bell-CHSH inequality \cite{Clauser:1969ny} 
\beq
{\rm Max}_{\{\vec{n}_i\}}  \Bigl| n_1^a C_{ab} (n_2^b+n_4^b) + n_3^aC_{ab}(n_2^b-n_4^b)
\Bigr| \le 2, \label{ch}
\eeq
for unit vectors $|\vec{n}_{i}|=1$ ($i=1,2,3,4$). This inequality is violated if the largest two of the three eigenvalues $\mu_3\le \mu_2\le \mu_1$ of the matrix $C^{\cal T}C$ (the symbol ${\cal T}$ denotes `transpose') satisfy \cite{Horodecki:1995nsk}
\beq
1<\mu_1+\mu_2\le 2.
\eeq
In the longitudinal case, we find 
$(C^L)^{\cal T}C^L= {\rm diag}(1,1,1)$
so that $\mu_1+\mu_2=2$ and the inequality is maximally violated, with the left hand side of (\ref{ch}) reaching $2\sqrt{2}$. Therefore, the $q\bar{q}$ pair always exhibits maximal entanglement and maximal Bell nonlocality both in single and multiple gluon exchanges. 

\begin{figure}[t]
%\vspace{-20mm}
    \centering
\includegraphics[width=1\linewidth]{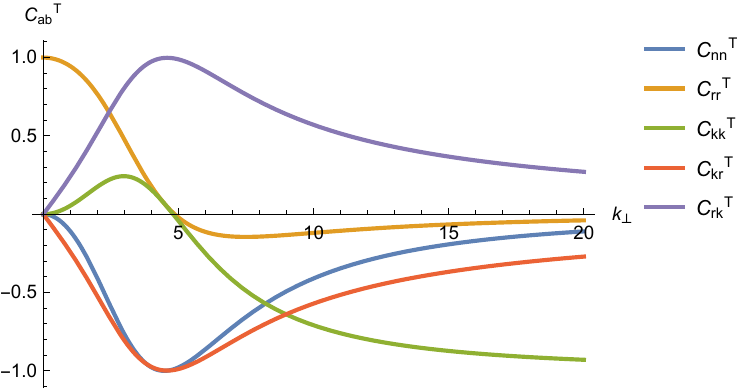}
    \caption{$C^T$-matrix at $z=1/2$ as a function of $k_\perp$ (in units of GeV) for the bottom quark $m=4.18$ GeV. We used the GBW model \cite{Golec-Biernat:1998zce} with $Q^2=9$ GeV$^2$ and $W=100$ GeV. }
    \label{s2}
\end{figure}

{\it Transverse photon}---The calculation is significantly more complicated for the transversely polarized virtual photon. Again we write the $\gamma^*_T+p$ cross section in the form (\ref{rep}) with $L\to T$. The same cross section in the photo-production limit $Q=0$ is relevant to proton-nucleus UPCs   \cite{Hagiwara:2017fye,ReinkePelicer:2018gyh,Goncalves:2019jdl,Linek:2023kga}.
The unpolarized cross section is well known 
\beq
A^T&=& \frac{2\alpha_{em}e_q^2}{N_c}((z^2+\bar{z}^2)k_\perp^2(T_1+T_2)^2 +m^2T_1^2)\,.
\eeq
The result for $C_{ab}^T$ is new but lengthy, and we present it in Appendix together with a brief outline of the calculation. 
In contrast to the longitudinal case (\ref{cij}), $C_{ab}^T$ is neither symmetric nor antisymmetric. (However,  $C_{kr}^T=-C_{rk}^T$ for  $z=\frac{1}{2}$.) Yet, we noticed  the following  nontrivial identities for generic values of $z$
\beq
(C_{rr}^T)^2+(C_{rk}^T)^2+(C_{kr}^T)^2+(C_{kk}^T)^2 -(C_{nn}^T)^2 = 1. \label{id1}
\eeq
\beq
C_{nn}^T = -C_{rr}^TC_{kk}^T+C_{rk}^TC_{kr}^T. \label{id2}
\eeq
In fact, the same identities hold in the longitudinal case (\ref{cij}) but in a more trivial manner. Fig.~\ref{s2} illustrates the behavior of the coefficients \( C_{ab}^T \) as a function of \( k_{\perp} \) at \( z = \frac{1}{2} \). Each curve shows a characteristic behavior around $k_\perp \approx \mu$.

\begin{figure}[t]
%\vspace{-20mm}
    \centering
    \includegraphics[width=1\linewidth]{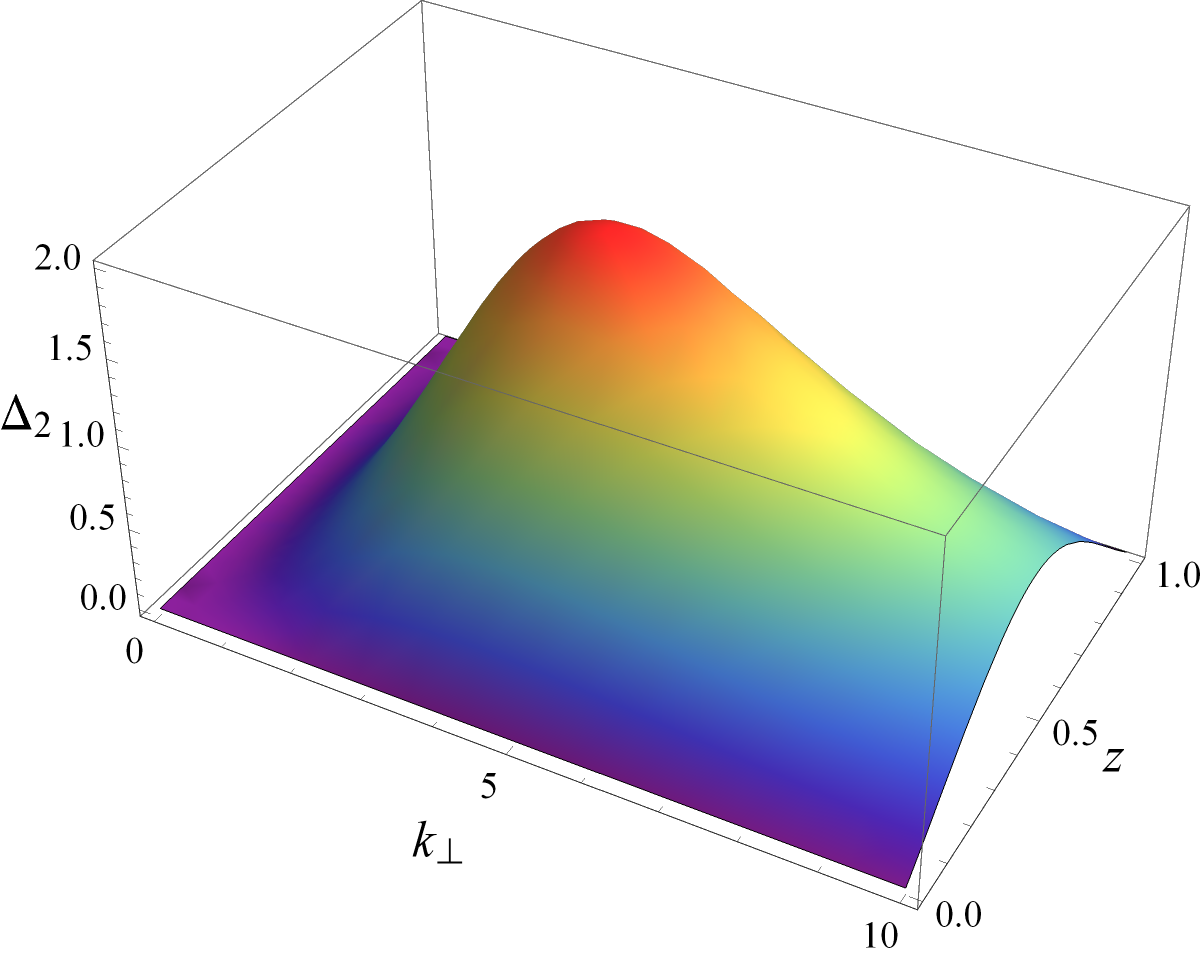}
    \caption{$\Delta_2$ in the $(z,k_\perp)$ plane at $Q^2=9$ GeV$^2$, $m=4.18$ GeV, and $W=100$ GeV.   }
    \label{s1}
\end{figure}

In contrast to the one-gluon exchange studied in  \cite{Qi:2025onf}, 
 $C^T_{ab}$ now depends on the structure of the target proton through the dipole T-matrix $T_{1,2}$. Remarkably, however, the criterion for  entanglement does not depend on $T_{1,2}$. 
Consider the following two quantities 
\beq
\begin{split}
\Delta_1= \sqrt{(C_{rr}^T-C_{kk}^T)^2+(C_{rk}^T+C_{kr}^T)^2}-1+C_{nn}^T, \\
\Delta_2= \sqrt{(C_{rr}^T+C_{kk}^T)^2+(C_{rk}^T-C_{kr}^T)^2}-1-C_{nn}^T. 
\end{split}
\eeq
According to the Peres-Horodecki criterion \cite{Peres:1996dw,Horodecki:1997vt}, if one of $\Delta$'s is nonnegative, the $q\bar{q}$ pair is entangled \cite{Afik:2020onf,Afik:2022kwm}. 
It follows from (\ref{id1}) and (\ref{id2}) that
\beq
\Delta_2=-\Delta_1 = -2C_{nn}^T\ge 0.
\eeq
Therefore, the $q\bar{q}$ pair is  always entangled except when $C_{nn}^T=0$, i.e., on the  kinematical end-points $z=0,1$, or exactly at the threshold $k_\perp=0$, or in the ultrarelativistic limit $\beta\to 1$  ($k_\perp\to \infty$). Again, this is in contrast to \cite{Qi:2025onf} (see also \cite{Afik:2025grr}) where there is a region of no entanglement (i.e., the system is separable), and entanglement grows stronger near the threshold or in the ultrarelativistic limit. 
$\Delta_2$ is plotted as a function of $z$ and $k_\perp$ in Fig.~\ref{s1}. 
For a fixed value of $k_\perp$, $\Delta_2$ has a peak at $z=\frac{1}{2}$ where 
\beq
C_{nn}^T=-\frac{k_\perp^2(T_1+T_2)^2}{k_\perp^2(T_1+T_2)^2+2m^2T_1^2}.
\eeq
This takes the minimum value $C_{nn}^T=-1$ and $\Delta_2=2$ takes the maximal value if and only if  $T_1=0$.  (Naively, $C_{nn}^T$ also approaches $-1$ in the ultrarelativistic limit $k_\perp\to \infty$. However, $C_{nn}^T$ vanishes in this limit (cf., Fig.~\ref{s2}) due to a cancellation between $T_1$ and $T_2$.) 
Interestingly, in heavy quark production,  $T_1$ does cross  zero. This can be shown model independently by expanding the integrand of  (\ref{t1t2}) in $|{\bm p}|$.  Using the color transparency condition  $\int d^2{\bm p}T({\bm p})=0$, 
%and neglecting the angular dependence of $T({\bm p})$, 
we  find 
\beq
T_1\approx 
 \frac{k_\perp^2-\mu^2}{(k_\perp^2+\mu^2)^3}\int d^2{\bm p}\,{\bm p}^2 T({\bm p}). \label{zero}
\eeq
Therefore, $T_1$ vanishes at $k_\perp \approx \mu=\sqrt{Q^2/4+m^2}$, and at this point the $C^T$-matrix takes the form 
\beq
C_{ab}^T\left(k_\perp \approx m, z=\frac{1}{2}\right)
\approx \begin{pmatrix} -1 & 0 & 0 \\ 0 & 0 & 1 \\ 
0 & -1 & 0 \end{pmatrix}, 
\eeq
where we have taken  the heavy quark mass limit.  
The corresponding density matrix can be written as 
\beq
\rho^T =\frac{1}{4}\left(1\otimes 1+C_{ab}^T\sigma^a\otimes \sigma^b\right) 
\approx 
|\Psi\rangle\langle \Psi| ,
\eeq
with 
\beq
|\Psi\rangle = \frac{1}{\sqrt{2}}\Bigl( |+\rangle^n_q|-\rangle^n_{\bar{q}} -i|-\rangle^n_q|+\rangle^n_{\bar{q}} \Bigr).
\label{pure}
\eeq
This is a maximally entangled state, with  the superscript $n$ indicating that the spin quantization axis is taken along the $+\hat{n}$ direction (not the $+\hat{k}$ direction).

Next, we test the CHSH inequality. 
With the help of (\ref{id1}) and (\ref{id2}), it is easy to check that the three eigenvalues of the matrix $(C^T)^{\cal T} C^T$ are 
\beq
\mu_1= 1, \qquad \mu_2=\mu_3 = (C_{nn}^T)^2.
\eeq
Therefore, the CHSH inequality is violated as long as $C_{nn}^T<0$,  which is  remarkably  the same condition as for nonvanishing entanglement. This coincidence can be attributed to the identities (\ref{id1}) and (\ref{id2}). In particular, when $C_{nn}^T=-1$ where $\mu_1+\mu_2=2$, the inequality is maximally violated. We therefore find that, even in the transversely polarized case, entanglement necessarily entails Bell nonlocality and vice versa. This is  in sharp contrast to the situation in the one-gluon exchange   \cite{Qi:2025onf}, or more generally, in typical two-qubit problems where Bell nonlocality is more nontrivial to realize than entanglement.  

\begin{figure}[t]
\vspace{-7mm}
    \centering
    \hspace{10mm}\includegraphics[width=1.7\linewidth]{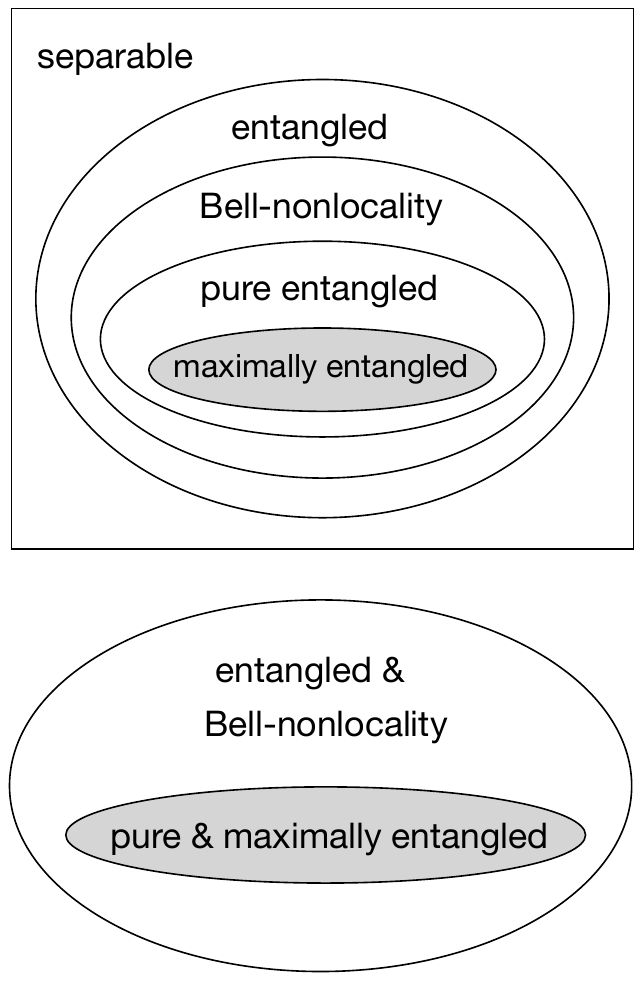}
    \vspace{-20mm}
    \caption{Top:  Venn diagram for generic two-qubit systems. Bottom: Venn diagram of a $q\bar{q}$ pair produced in $\gamma^*_T+{\mathbb P}\to q+\bar{q}$. The `separable' region is the boundary of the entangled region.  If the photon is longitudinally polarized, only the shaded (`pure \& maximal') region   is present.  }
    \label{venn}
\end{figure}

According to Gisin's theorem \cite{Gisin:1991vpb},  for any two-qubit system that is a pure entangled state, the Bell-CHSH inequality is violated. Indeed, in the longitudinally polarized case, the density matrix satisfies the pure state condition $(\rho^L)^2=\rho^L$. However, the transverse density matrix $\rho^T$ does not in general represents a pure state since $(\rho^T)^2\neq \rho^T$. 
The equality holds only if $C_{nn}=-1$, in which case the state becomes pure and maximally entangled (\ref{pure}). This  is depicted in  Fig.~\ref{venn} (bottom), to be compared with the nesting structure for generic two-qubit systems (top). %The peculiarity of the present system is obvious.  

Finally, we briefly discuss prospects for measurements  using UPCs as an example, while leaving detailed feasibility studies for future work. 
In order to access spin information, the following  channels have been  identified as promising \cite{Kats:2023zxb} 
\beq
&& b\xrightarrow{7.0\%}  \Lambda_b\xrightarrow{11\%} X^+_c+\mu^-+\bar{\nu}_\mu, \label{1}
\nn
&& c\xrightarrow{6.4\%}\Lambda_c^+ \xrightarrow{3.5\%} \Lambda+\mu^++\nu_\mu, \label{frag}\\
&& s\xrightarrow{2.8\%} \Lambda \xrightarrow{64\%} p+\pi^-, \notag
\eeq
and their charge-conjugate counterparts for $\bar{b},\bar{c},\bar{s}$. The numbers above the first arrow are the fragmentation probabilities.
The spin of the  (anti)quark is largely inherited by the daughter (anti)baryon \cite{Galanti:2015pqa}.  Similarly, especially for heavy quarks, roughly  $z^{\rm quark}\approx z^{\rm hadron}$ and $k_\perp^{\rm quark}\approx k_\perp^{\rm hadron}$, to first approximation. The numbers above the second arrow are the branching ratios. 
Once these events have been identified and the baryon and antibaryon momenta have been reconstructed from their  decay products \cite{Galanti:2015pqa,Kats:2023zxb,Afik:2025grr}, one goes to their  respective rest frames by first moving to the pair's CM frame and then boosting along the $\pm \hat{k}$ directions. In each rest frame,  the angular distribution of one of the decay particles can be measured.
In order to access the component $C_{nn}^T$, one needs the polar angle distribution  of each decay particle with respect to the  $\hat{n}$ axis in the form
\beq
\frac{1}{\sigma}\frac{d\sigma^{\rm UPC}}{d\cos\theta_+d\cos\theta_-}  = \frac{1+\alpha_+\alpha_- r_T^2 \cos\theta_+\cos\theta_-C_{nn}^T}{4},
\eeq
where $\alpha_+$ and $\alpha_-=-\alpha_+$ are the spin analyzing power of the measured particle and antiparticle, and $r_T$ is the transverse polarization  retention factor \cite{Galanti:2015pqa}.  Equivalently, 
\beq
C_{nn}^T=\frac{9}{\alpha_+\alpha_-r_T^2}\langle \cos\theta_+\cos\theta_-\rangle.
\eeq
Since $C_{nn}^T\le 0$, we predict that $\langle \cos\theta_+\cos\theta_-\rangle \ge 0$, and this signifies both entanglement and Bell-nonlocality, although the effect of hadronization should be carefully investigated.  To give a rough idea about  the expected yield, we consider  UPCs with the lead nucleus ($A=208,Z=82$) at the LHC energy $\sqrt{s_{NN}}=8.1$ TeV. According to   \cite{Linek:2023kga}, $d\sigma_{c\bar{c}}^{\rm UPC}/dk_\perp \sim {\cal O}(100 \,\mu{\rm b}/{\rm GeV})$ at $k_\perp \sim m_c$. We have confirmed this result and moreover predict that  $ d\sigma_{b\bar{b}}^{\rm UPC}/dk_\perp\sim {\cal O}(10 \, {\rm nb}/{\rm GeV})$ at $k_\perp\sim m_b$ and $ d\sigma_{s\bar{s}}^{\rm UPC}/dk_\perp\sim {\cal O}(100\,  \mu{\rm b}/{\rm GeV})$. (Here we quote the value at $k_\perp \gtrsim  1$ GeV since, for the $s$-quark, the region $k_\perp\sim m_s$ is nonperturbative.)  The fragmentation and branching probabilities (\ref{frag}) bring in a suppression  factor of order $10^{-4}$ for the $s\bar{s}$ pair and $10^{-6}$ for the $c\bar{c},b\bar{b}$ pairs. Further considering $\alpha_\pm,r_T \sim {\cal O}(1)$  \cite{Galanti:2015pqa}, we find the signal cross sections of order 10nb, 100pb, 0.01fb for the $s\bar{s},c\bar{c},b\bar{b}$ channels, respectively. Assuming  $1\sim 10$pb$^{-1}$ of proton-lead collision data at the future high-luminosity  LHC experiment \cite{dEnterria:2025jgm}, our  rough estimates indicate that the $s\bar{s}$-channel is feasible, while the $c\bar{c}$-channel is marginal and the $b\bar{b}$-channel is impractical at the LHC. In future, the discussion will be extended to light quarks \cite{Hatta:2025obw} that have larger cross sections. Also,  other methods to measure the spin correlation without relying  on weak decays should be explored \cite{Cheng:2025cuv}.  

{\it Conclusions}---We have calculated the spin density matrix of a $q\bar{q}$ pair  originating from the Pomeron, the color singlet gluonic exchange in the $t$-channel that governs  QCD amplitudes at high energy.  In the longitudinally polarized case, we find maximal entantlement and maximal violation of the Bell-CHSH inequality. This agrees with  the result from the one-gluon exchange in \cite{Qi:2025onf}, although such an agreement is nontrivial. More strikingly, the transversely polarized photon in DIS or UPCs always leads to entanglement and Bell-nonlocality in the produced $q\bar{q}$ pair. In other words, in the present system, entanglement is a necessary {\it and sufficient} condition for Bell-nonlocality. This remarkable feature of the Pomeron, revealed in this work for the first time, is quite peculiar from the viewpoint of quantum information science (see Fig.~\ref{venn}) and certainly  deserves further investigation.

\section*{Acknowledgments}

We thank Bowen Xiao for stimulating discussions. 
M.~F. thanks the EIC theory institute of Brookhaven National Laboratory, where this work was initiated, for hospitality. Y.~H. is supported by the U.S. Department
of Energy under Contract No. DE-SC0012704, by LDRD funds from Brookhaven Science Associates, and also by the framework of the Saturated Glue (SURGE) Topical Theory Collaboration. The work of M.~F. is
supported by the ULAM fellowship program of NAWA No. BNI/ULM/2024/1/00065 “Color glass condensate effective theory beyond the eikonal approximation”.

\begin{widetext}
\appendix

\section{Spin density matrix element for the transversely polarized photon}
In this appendix we sketch the calculation of the spin density matrix in the transverse case. Averaging over the two photon polarizations, we write the cross section as   
\beq
\left. \frac{d \sigma^{ T}_{\alpha \alpha' \beta \beta'}}{d z  d^2 \boldsymbol{k}d^2{\bm \Delta} }\right|_{ {\bm \Delta}\approx 0}  &=&  \frac{\alpha_{em}e_q^2}{8z\bar{z}N_c(q^+)^2} \int   \frac{ d^2 \boldsymbol{p}T \left( {\bm p} \right) }{(\boldsymbol{k}-\boldsymbol{p})^2+\mu^2} \int  \frac{ d^2 \boldsymbol{p}' T \left( {\bm p}' \right) }{(\boldsymbol{k}-\boldsymbol{p}')^2+\mu^2}  \nonumber \\ &&\times \bar{v}_{\beta'} ( \tilde{k}' ) \left( (1-2z) (k^i-p^i) \gamma^+ - m \gamma^+ \gamma^i + i\epsilon^{ij}(k_{j}-p_{j})\gamma^5 \gamma^+ \right)  u_{\beta}( \tilde{k}) \nonumber \\ &&\times \bar{u}_{\alpha} ( \tilde{k} ) \left( (1-2z) (k^i-p'^i) \gamma^+ - m \gamma^i \gamma^+ + i\epsilon^{ij}(k_{j}-p'_{ j})\gamma^+\gamma^5 \right)  v_{\alpha'}( \tilde{k}'), \label{mi}
 \nn
&=&\frac{\alpha_{em}e_q^2}{8z\bar{z}N_c(q^+)^2} \bar{v}_{\beta'} ( \tilde{k}' ) \left[ \left((1-2z) k^i \gamma^+ +i\epsilon^{ij}k_{ j}\gamma^5\gamma^+\right) (T_1+T_2) - mT_1 \gamma^+ \gamma^i \right]  u_{\beta}( \tilde{k}) \nonumber \\ &&\times \bar{u}_{\alpha} ( \tilde{k} ) \left[ \left((1-2z) k^i \gamma^+ +i\epsilon^{ij}k_{ j}\gamma^+\gamma^5 \right)(T_1+T_2) - mT_1 \gamma^i \gamma^+  \right]  v_{\alpha'}( \tilde{k}').  \label{tsigma}
\eeq 
where $i,j=1,2$ and $\epsilon^{12}=-\epsilon^{21}=1$. 
We first observe that the matrix elements   (\ref{tsigma}) do not depend on the azimuthal angle of ${\bm k}$. Therefore,  without loss of generality, we may take ${\bm k}=(k_\perp,0)$ or $\phi=0$.
Next we use the conversion rules which  simplify when $\phi=0$ 
\beq
\gamma^+ = \frac{1}{\sqrt{2}}\left(\gamma^0 +\sin \theta \gamma^r+\cos\theta \gamma^k\right), \qquad \gamma^1= -\cos\theta \gamma^r + \sin \theta \gamma^k, \qquad 
\gamma^2=  \gamma^n .
\eeq 
In the $(\hat{n},\hat{r},\hat{k})$ system, spinors in  the helicity basis read  %\cite{Jacob:1959at,Haber:1994pe}  
\beq
u_{\alpha}(k)=\begin{pmatrix} \sqrt{k\cdot \sigma}\xi_\alpha \\ \sqrt{k\cdot \bar{\sigma}}\xi_\alpha \end{pmatrix}, 
\qquad v_{\alpha'}(k')= \begin{pmatrix} \sqrt{k'\cdot \sigma}\tilde{\eta}_{-\alpha'} \\ -\sqrt{k'\cdot \bar{\sigma}}\tilde{\eta}_{-\alpha'} \end{pmatrix} = \begin{pmatrix} \sqrt{k\cdot \bar{\sigma}}\tilde{\eta}_{-\alpha'} \\ -\sqrt{k\cdot \sigma}\tilde{\eta}_{-\alpha'} \end{pmatrix}, \label{heli}
\eeq
with ($\sigma^k=\sigma^3$)
\beq
\sqrt{k\cdot \sigma}
= \sqrt{\frac{M}{8}}\left( \sqrt{1+\beta}(1-\sigma^k)+\sqrt{1-\beta}(1+\sigma^k)\right), \qquad  
\sqrt{k\cdot \bar{\sigma}} = \sqrt{\frac{M}{8}}\left( \sqrt{1-\beta}(1-\sigma^k)+\sqrt{1+\beta}(1+\sigma^k)\right).
\eeq
Here, $\alpha,\alpha'=\pm$ refers to the helicity (angular momentum projection $\pm \frac{1}{2}$ along the direction of motion). For the antiquark moving in the $-\hat{k}$ direction, the `flipped spinors' 
$\tilde{\eta}_{-\alpha'}=-i\sigma^2 (\eta_{\alpha'})^*$  \cite{Peskin:1995ev} 
are used. 
We then exploit the fact that any 2 $\times$ 2 matrix can be expanded in the basis $\{ 1, \sigma^n, \sigma^r, \sigma^k \}$ 
\beq
\xi_\beta \xi^\dagger_{\alpha} &=& \frac{1}{2} \left( \xi^\dagger_\alpha \xi_\beta + (\xi^\dagger_\alpha \sigma^n \xi_\beta) \sigma^n + (\xi^\dagger_\alpha \sigma^r \xi_\beta) \sigma^r +(\xi^\dagger_\alpha \sigma^k \xi_\beta) \sigma^k \right) \; , \nn 
\tilde{\eta}_{-\alpha'} \tilde{\eta}^\dagger_{-\beta'} &=& \frac{1}{2} \left( \tilde{\eta}^\dagger_{-\beta'} \tilde{\eta}_{-\alpha'} + (\tilde{\eta}^\dagger_{-\beta'} \sigma^n \tilde{\eta}_{-\alpha'}) \sigma^n + (\tilde{\eta}^\dagger_{-\beta'} \sigma^r \tilde{\eta}_{-\alpha'}) \sigma^r +(\tilde{\eta}^\dagger_{-\beta'} \sigma^k \tilde{\eta}_{-\alpha'}) \sigma^k \right) \; . \label{xixi}
\eeq
Finally we use the formula $
\tilde{\eta}_{-\beta'}^\dagger \vec{\sigma}\tilde{\eta}_{-\alpha'} =
-\eta^\dagger_{\alpha'}\vec{\sigma}\eta_{\beta'}$
to recast the density matrix in the form (\ref{rep}).   
After tedious calculations, we find $
C_{ab}^T=\widetilde{C}^T_{ab}/A^T$
where 
\beq
\widetilde{C}_{nn}^{T} = - \frac{4\alpha_{em}e_q^2}{N_c} z\bar{z}k_\perp^2(T_1+T_2)^2,   
\eeq
\beq
\widetilde{C}_{rr}^{T} &=& \frac{\alpha_{em}e_q^2}{2z\bar{z}N_c}  \bigg[ 2 T_1 (T_1+T_2) k_\perp m \sqrt{1-\beta ^2}  
   \sin  \theta  (\beta + (1-2 z) \cos \theta)+ T_1^2 m^2 \left( 1 - \beta ^2 \right)  \sin^2\theta \nn 
   && - (T_1+T_2)^2 k_\perp^2  \bigg( \left(1-\beta ^2\right) \sin ^2\theta + 2 z (1-z) \left( \left(2-\beta ^2\right)  \cos ^2\theta  -1 \right) \bigg) \bigg],
\eeq
\beq
\widetilde{C}_{kk}^{T} &=& \frac{\alpha_{em}e_q^2}{2z\bar{z}N_c} \bigg[ 2 T_1 (T_1+T_2) k_\perp m \sqrt{1-\beta ^2} \sin \theta  (\beta - (1 -2 z) \cos \theta )+ T_1^2 m^2 \left(\cos ^2\theta -\beta ^2\right) \nn && 
+ (T_1+T_2)^2 k_\perp^2 \bigg(\beta ^2 - 2 z (1-z)  + \cos ^2\theta  \big(2 z(1-z) \left(2 - \beta^2\right)  - 1 \big)  \bigg) \bigg] ,
\eeq

\beq
\widetilde{C}_{r k}^{T} &=& \frac{\alpha_{em}e_q^2}{2z\bar{z}N_c}  \bigg[ 2 T_1 (T_1+T_2) k_\perp m
   \left( z -\beta ^2+\cos ^2\theta \left( 1 - \left(2- \beta ^2\right)
   z \right) \right) \nonumber \\ && + \sqrt{1-\beta ^2} \sin \theta  \left( T_1^2
   m^2 (\cos \theta -\beta )+ (T_1 + T_2)^2 k_\perp^2 \left(\beta
   -(1-2 z)^2 \cos \theta \right)\right) \bigg] \; ,
\eeq

\beq
\widetilde{C}_{k r}^{T} &=& \frac{\alpha_{em}e_q^2}{2z\bar{z}N_c} \bigg[ 2 T_1 (T_1+T_2) k_\perp m
   \left( z-\left(1 - \beta ^2 \right) \sin ^2\theta -\left(2-\beta ^2\right) z
   \cos ^2\theta  \right) \nonumber \\ && + \sqrt{1-\beta ^2} \sin \theta 
   \left( T_1^2 m^2 ( \beta + \cos \theta) - (T_1+T_2)^2 k_\perp^2 \left( \beta +(1-2 z)^2 \cos \theta \right) \right) \bigg] \; .
\eeq
One may eliminate $\beta,\theta$ in favor of $z,k_\perp$ using (\ref{beta}) and (\ref{theta}).  While the result is not more concise, it shows that the apparent pole $1/z\bar{z}$ is actually absent. 

\end{widetext}

\bibliography{ref}

\end{document}